\documentclass[12pt,a4paper]{article}
\usepackage[latin1]{inputenc}
\usepackage[english]{babel}

\usepackage{amsfonts}
\usepackage{amsmath}
\usepackage{amsthm} % for proof environment
\usepackage{bm} % for bm
\usepackage[round]{natbib}
\usepackage{graphicx}
\usepackage{caption}
\usepackage{subcaption}
\usepackage[parfill]{parskip}
\usepackage{epstopdf}
\usepackage[space]{grffile}
\usepackage{appendix}
\usepackage{booktabs}
\usepackage{blkarray}
\usepackage{verbatim}

\newcommand{\dd}{{\mathrm d}}

\newcommand{\R}{{\mathbb R}}

\newcommand{\Lcal}{{\mathcal L}}

\makeatletter
\def\thm@space@setup{%
  \thm@preskip=\parskip \thm@postskip=0pt
}
\makeatother

\newtheorem{proposition}{Proposition}[section]
\newtheorem{lemma}[proposition]{Lemma}
\newtheorem{theorem}[proposition]{Theorem}

\newtheorem{remark}[proposition]{Remark}

\title{ Exact Smooth Term-Structure Estimation\footnote{
We thank Leif Andersen, Jeroen Kerkhof, Fabio Mercurio, and two anonymous referees for comments. The research leading to these results has received funding from the European Research Council under the European Union's Seventh Framework Programme (FP/2007-2013) / ERC Grant Agreement n.~307465-POLYTE.}}
\author{Damir Filipovi\'c\footnote{EPFL and Swiss Finance Institute. Email: damir.filipovic@epfl.ch} \quad \quad Sander Willems\footnote{EPFL and Swiss Finance Institute. Email: sander.willems@epfl.ch}}
\date{February 12, 2018\\ forthcoming in \textit{SIAM Journal on Financial Mathematics}}
\providecommand{\noopsort}[1]{}

\begin{document}
\maketitle
\begin{abstract}
We present a non-parametric method to estimate the discount curve from market quotes based on the Moore--Penrose pseudoinverse. The discount curve reproduces the market quotes perfectly, has maximal smoothness, and is given in closed-form. The method is easy to implement and requires only basic linear algebra operations. We provide a full theoretical framework as well as several practical applications.
\end{abstract}

\medskip



\section{Introduction}
In financial models it is often assumed that we can observe an initial term-structure of zero-coupon bond prices for the continuum of maturities, also known as the discount curve. In practice, however, zero-coupon bonds are rarely traded and the discount curve has to be derived from prices of actively traded fixed-income instruments such as coupon bonds, interest rate swaps or futures. Since the discount curve is an infinite-dimensional object, we need an interpolation method to complete the information obtained from the finite number of observed market instruments. Broadly speaking we can divide term-structure estimation methods in two categories: parametric methods and non-parametric methods.

Parametric methods impose a particular functional form for (parts of) the discount curve and calibrate the parameters by minimizing the pricing error. Examples of single-piece functions that are defined over the entire maturity domain include the seminal work of \citet{nelson} and \citet{svensson}. These are typically low-dimensional parametric forms and are preferred for more qualitative studies where the general shape of the curve is more important than the exact values (e.g.\ monetary policy in central banks). Single-piece functional forms are however too restrictive for institutions involved in trading as they prefer to have a discount curve that perfectly reproduces market quotes in order to mark to market their books within a single arbitrage-free valuation framework. Rather than specifying a single function for the entire maturity spectrum, polynomial spline methods impose a piecewise polynomial specification. The first application in term-structure estimation goes back to \citet{mcculloch1971measuring,mcculloch1975tax} where quadratic and cubic splines are fitted directly to the discount function using ordinary least squares regressions. \cite{steeley1991estimating} proposed the use of B-splines to overcome ill-conditioned matrices encountered in \citet{mcculloch1971measuring,mcculloch1975tax}. We refer to \citet{hagan2006interpolation} for a survey of several other spline based algorithms. A close fit to market data can be achieved by increasing the number of knot points in the spline. The choice of both the number and the positions of the knot points remains, however, completely ad hoc.

A second class of estimation methods are the non-parametric approaches. Instead of imposing a particular functional form on (a transformation of) the discount curve, these methods minimize a norm that is related to the smoothness and goodness-of-fit of the curve. Several definitions of smoothness have been considered in the literature. \cite{delbaen1992estimation} and \cite{frishling1996fitting} minimize the integrated squared first derivative of the forward curve, arguing that forward rates over various harizons should not vary too much. \cite{adams1994fitting} and \cite{lim2002computing}, among others, use the integrated squared second order derivative of the forward curve as a measure of smoothness. Both approaches lead to polynomial splines for the optimal forward curve. \cite{manzano2004positive}, \cite{andersen2007discount}, and \cite{kwon2002general} consider combinations of these two measures and show that this results in so called hyperbolic or tension splines. All of the above papers smooth a transformation of the discount curve (typically the forward curve) and numerical routines have to be invoked to solve for the optimal curve. Exceptions are the works of \cite{delbaen1992estimation} and \cite{adams1994fitting} for the special case where the set of benchmark instruments consists solely of zero-coupon bonds. In reality, however, zero-coupon bonds are hardly ever liquidly traded and the discount curve has to be estimated based on coupon bearing bonds or swaps rates.

In this paper we introduce an easy to use non-parametric method based on the Moore--Penrose pseudoinverse. We search in an infinite-dimensional Hilbert function space for a discount curve that has minimal norm and exactly prices a benchmark set of linear fixed-income instruments, e.g.\ FRAs, swaps, or coupon bonds. The norm is related to the integrated squared second derivative of the discount curve. The optimal discount curve is given by a cubic spline with knot points positioned exactly at the cashflow dates of the benchmark instruments. Because we directly smooth the discount curve function, the optimal curve is given in closed form and requires only simple linear algebra calculations. The methodology in this paper closely resembles that of \cite{lorimier}, \cite{adams1994fitting}, \cite{tanggaard1997nonparametric}, \cite{andersen2007discount}, and others, however they all focus on estimating transformations of the discount curve (e.g., the forward curve). To the best of our knowledge, this paper is the first to present a fully worked out treatment of non-parametrically estimating the discount curve itself. We argue that our method is a valuable and easy to use alternative to more complex numerical algorithms to find a smooth discount curve.

Our method is designed to exactly reproduce the prices of benchmark instruments. This is common practice when the benchmark instruments are liquid Libor related instruments (e.g., swaps). The prices are typically taken to be the mid-prices. When building discount curves using coupon bonds, bid-ask ranges are often wider and a discount curve is in principle allowed to produce any price that lies within this range. We show how to optimally pick the prices within bid-ask ranges such that the smoothness of the discount curve is maximally increased. In case the benchmark instruments are coupon bonds, we show that this reduces to solving a convex quadratic programming problem with linear inequality constraints. 

As highlighted initially by \cite{vasicek1982term} and \cite{shea1984pitfalls}, fitting a polynomial spline directly to the discount curve need not lead to a positive nor a monotonically non-increasing discount curve. \cite{barzanti1998note} use tension splines where the tension in the spline is increased manually until problematic behaviour is avoided. \cite{chiu2008approximating}, \cite{laurini2010constrained}, and \cite{fengler2015simple}, among others, impose shape constraints on the B-splines used to represent the discount function. The discount curve produced by our method is not guaranteed to be positive or monotonic non-increasing, however we did not find this to be a problem in the numerical examples we have explored. We develop a finite-dimensional counterpart of our method for which positivity and monotonicity constraints can easily be incorporated by numerically solving a convex quadratic programming problem with linear inequality constraints. 

This paper is structured as follows. Section \ref{secEP} casts the term-structure estimation problem and shortly reviews the steps taken in a traditional bootstrap. Section \ref{sec:theory} presents the theory behind our proposed method. In Section \ref{sec:sensitivities} we discuss the sensitivity of the optimal discount curve with respect to the input prices. In particular this section shows how to optimally choose prices from a bid-ask range. Section \ref{sec:examples} illustrates our method with market data. Section \ref{sec:finite_dim} contains a finite-dimensional equivalent of our method. Section \ref{sec:conclusion} concludes. The appendix contains all proofs.

\section{Estimation problem}\label{secEP}
Suppose today is time 0 and denote by $p=(p_1,\ldots, p_n)^\top$ the observed prices of $n$ fixed-income instruments. Denote by $0\leq x_1<\cdots < x_N$ the union of all the cashflow dates of these instruments and call $C=(c_{ij})$ the corresponding $n\times N$ cashflow matrix.\footnote{By cashflow dates we mean every date that is relevant for the pricing of the instrument.} If instrument $i$ does not have a cashflow at time $x_j$, then we simply set $c_{ij}=0$.
The information contained in these $n$ instruments about the discount curve can be summarized by a linear system as follows:
\begin{equation}
Cd=p,
\label{eq:pricingEquation}
\end{equation}
with $d=(g(x_1),\ldots,g(x_N))^\top$ and where $g(x)$ denotes the price  of a zero-coupon bond maturing in $x$ years. If $C$ were an invertible square matrix, then there would exist a unique solution to this system: $d=C^{-1}p$. In reality, however, we typically have many more cashflow dates than instruments ($n\ll N$) that we can use for the estimation. In other words, the linear system $Cd=p$ is under-determined and there exist many discount vectors $d$ that satisfy the relation in \eqref{eq:pricingEquation}.\footnote{Assuming the system is not inconsistent, i.e.\ instruments that can be replicated as a linear combination of other instruments must have the same price (no-arbitrage).} The first problem that arises is therefore which of the admissible discount vectors should be chosen. Second, once we have chosen a particular admissible vector $d$, we still face an interpolation problem to find $g(x)$ for $x\in (0,x_1)$ and $x\in (x_{i-1},x_i)$, $i=2,\ldots,N$.

Bootstrapping is a common practice among trading desks to construct a discount curve from a limited number of carefully selected liquid market instruments such that the resulting curve perfectly reproduces the prices of the instruments used in the estimation.
There is no unique bootstrapping method and it is likely that there are at least as many methods as there are trading desks in the world. In this section we give a very brief description of some methods, for a more detailed overview of the most popular (single-curve) bootstrapping methods used in practice we refer to \citet{hagan2006interpolation,hagan2008methods}.
In general, one can a priori impose an explicit parametric form for the discount curve: $g(x)=g(x;z)$ for some parameter $z$ with dimension less than or equal to the number of observed instruments $n$. The pricing system~\eqref{eq:pricingEquation} then becomes a system of possibly nonlinear equations in $z$:
\[C (g(x_1;z),\ldots,g(x_N;z))^\top=p.\]
Assuming that the gradients $\nabla_z g(x_j,z)$, $j=1,\dots,N$, are linearly independent, the inverse mapping theorem asserts that this system is no longer (locally) under-determined with respect to the parameter $z$. If it admits a solution $z^\ast$ then it is (locally around $z^\ast$) unique. The choice of a suitable parametric form $g(x;z)$ is however not straightforward. One possibility is to choose a polynomial of degree $n-1$, also known as the Lagrange polynomial. Although this function is very smooth and flexible enough to satisfy $n$ constraints, it demonstrates strong oscillatory behavior. A standard solution to this so called `roller coaster' effect is to describe the discount curve by splines, i.e.\ piece-wise low-dimensional polynomials. There are many different ways to specify the functional form of a spline and the position of the knot points, see for example \citet{mcculloch1971measuring,mcculloch1975tax}, \citet{steeley1991estimating} and \citet{adams2001smooth}. It is important to note that in all these cases the spline solution is imposed \emph{a priori} and both the number and the position of the knot points are chosen manually. The method we present in the next section also produces a spline. However, it remains fully non-parametric in the sense that the spline is the outcome of a proper optimization problem which determines the optimal number and position of the knot points.

\section{Pseudoinverse on Hilbert spaces}\label{sec:theory}
Instead of first finding a discount vector $d=(g(x_1),\ldots,g(x_N))^\top$ satisfying~\eqref{eq:pricingEquation} and in a second step interpolating these discount factors to a continuous discount curve, we now directly search for a discount curve in a convenient Hilbert function space that is optimal in the sense of having minimal norm. The optimal discount curve is explicitly calculated through the pseudoinverse of a continuous linear map.

We fix a finite time to maturity horizon $\bar\tau$ large enough to contain all cashflow dates. We define the Hilbert space of discount curves $H$ which consists of real functions $g\colon [0,\bar{\tau}]\to \R$ with absolutely continuous first derivative and norm given by
\begin{equation}
\| g\|_{H}^2=\langle g,g\rangle_{H} = g(0)^2 +g'(0)^2+ \int_0^{\bar{\tau}} g''(x)^2\,\mathrm{d}x.\label{eq:norm2}
\end{equation}

This norm serves as a measure of smoothness for the discount curve, which approximately captures the `flatness' of the corresponding forward curve. The forward rate is the rate one can lock in today on a riskless loan over a future time period. Unless there are specific reasons to believe otherwise, the forward rate should not fluctuate too much from one period to the next. 
If we denote by $F(x,y)$ the simple forward rate over a future time period $[x,y]$, $0<x<y$, then we have for small $h>0$:
\begin{align*}
g''(x)&\approx  \frac{g(x-h)-2g(x)+g(x+h)}{h^2}\\
&= \frac{g(x)}{h}\left(F(x-h,x)+\frac{1}{h}\left(\frac{1}{1+hF(x,x+h)}-1\right)\right)\\
&\approx \frac{g(x)}{h}\Big(F(x-h,x)-F(x,x+h)\Big).
\end{align*}
Hence, by minimizing the curvature of the discount curve, we are approximately minimizing the difference between subsequent simple forward rates.

For any $\tau\in[0,\bar{\tau}]$ we now define the linear functional $\Phi_\tau: H\to \R$ which evaluates the discount curve at $\tau$:
\begin{equation}\label{Phidef}
 \Phi_\tau(g)=g(\tau).
\end{equation}
By the Riesz representation theorem there exists a unique element $\phi_\tau\in H$ such that for any $g\in H$ we have
\[
\Phi_\tau(g)=\langle \phi_\tau, g \rangle_{H}.
\]
The following lemma gives an explicit expression for this element.
\begin{lemma}\label{lemma1}
The linear functional $\Phi_\tau$ on $H$ can be uniquely represented by the element $\phi_\tau\in H$ given by
\[
\phi_\tau(x)=1-\frac{1}{6}(x\wedge \tau)^3+\frac{1}{2}x\tau\left(2+x\wedge \tau\right),
\]
where we write $x\wedge \tau:=\min(x,\tau)$.
\end{lemma}

Let us define the linear map $M\colon H \to \R^n$ by
\begin{equation}\label{Mdef}
M g= C(\Phi_{x_1}(g),\ldots,\Phi_{x_N}(g))^\top,
\end{equation}
where $C$ is just as before the $n\times N$ cashflow matrix. We henceforth assume that $C$ has full rank. This is without loss of generality. Indeed, if $C$ did not have full rank, we would be including redundant instruments in our estimation as they can be replicated by linear combinations of other instruments. For example, two coupon bonds with different principal but otherwise identical characteristics impose the same constraints on the discount curve (assuming their prices are consistent with the law of one price).

We now find the discount curve with minimal $H$-norm that matches all benchmark quotes. That is, we solve the following infinite-dimensional optimization problem:
\begin{equation}
\begin{array}{cc}
\displaystyle
\min_{g\in H} & \frac{1}{2}\|g\|_{H}^2\\
\text{s.t.}			 & M g=p.
\end{array}
\label{optim2}
\end{equation}

The solution of \eqref{optim2} is an explicit piecewise cubic function, as shown in the following theorem:
\begin{theorem}\label{thm2}
There exists a unique solution $g^\ast\in H$ to the optimization problem~\eqref{optim2} and it is given as
\begin{equation}
 g^\ast(x) =(M^+p)(x)=z^\top \phi(x),
 %\sum_{j=1}^N z_j\,\phi_{x_j}(x),
 \label{eq:second_order_curve}
\end{equation}
where $M^+\colon \R^n\to H$ denotes the Moore--Penrose pseudoinverse of $M$, $z=C^\top\left(C A C^\top\right)^{-1} p$, $\phi(x)=(\phi_{x_1}(x),\ldots,\phi_{x_N}(x))^\top$, and $A$ is the positive definite $N\times N$-matrix with components $A_{ij}=\phi_{x_i}(x_j)=\phi_{x_j}(x_i)$.
\end{theorem}

We have therefore explicitly constructed the discount curve $x\mapsto g^\ast(x)$ that exactly replicates the prices $p$ of the instruments with cashflows $C$ and moreover it is the smoothest curve to do so among all real functions with absolutely continuous first derivative in the sense that it minimizes the norm in \eqref{eq:norm2}. The corresponding instantaneous forward curve $f^\ast\colon [0,\bar{\tau}]\to\R$ is given explicitly by:
\begin{align*}
f^\ast(x)&=-\frac{\mathrm{d}}{\mathrm{d}x}\ln(g^\ast(x))=-\frac{z^\top\phi'(x)}{z^\top\phi(x)},
%-\frac{\sum_{j=1}^N z_j\,\phi'_{x_j}(x)}{\sum_{j=1}^N z_j\,\phi_{x_j}(x)}
\end{align*}
where $\phi'(x)=(\phi'_{x_1}(x),\ldots,\phi'_{x_N}(x))^\top$ and $\phi'_{x_j}(x)=x_j-\frac{1}{2}(x\wedge x_j)^2 +x_j(x\wedge x_j)$.

\begin{remark}\label{remark:dynamic_arbitrage}
\cite{gourieroux2013linear} characterize dynamic term-structure models in which the zero-coupon bond prices are of the form $P(t,T)=z_t^\top a(T-t)$, where $z_t$ is a set of $N\ge 1$ linearly independent stochastic factors and $a:\R_+\to \R^N$ is a deterministic function. They show that the absence of arbitrage opportunity for a self-financed portfolio of zero-coupon bonds implies that there must exist a matrix $M$ such that $a'(\tau)=M a(\tau)$. The function $\phi=(\phi_{x_1},\ldots,\phi_{x_N})^\top$ does not satisfy this requirement and hence \eqref{eq:second_order_curve} is not consistent with a dynamic term-structure model. Similarly, the \cite{nelson} and \cite{svensson} specifications are also not arbitrage-free in a dynamic sense, see e.g.\ \cite{filipovic2000exponential}.
\end{remark}

The terms $g(0)^2$ and $g'(0)^2$ are included in \eqref{eq:norm2} to guarantee the definiteness of the norm. Since the discount curve must start at face value, we henceforth impose $g(0)=1$ by setting $x_1=0$, $p_1=1$, $c_{11}=1$, and $c_{1j}=c_{i1}=0$ for all $i,j\neq 1$. Hence, minimizing $g(0)^2$ does not influence the optimal curve since it is fixed in the constraints. 

The term $g'(0)^2$ leads to a minimization of the instantaneous short rate. If this is not desirable, we can easily fix the short rate to an exogenously specified value $r\in\R$. Indeed, note that $\psi(x)=x$ is the Riesz representation of the linear functional $\Psi(g)=g'(0)$ in $H$. We now add $\Psi(g)=-r$ as an additional constraint in \eqref{optim2} and find (analogously as in the proof of Theorem \ref{thm2}) the following unique solution: 
\begin{equation}
g^\ast(x)=\tilde{z}^\top \tilde{\phi}(x),
%\sum_{j=1}^N \tilde{z}_j\,\phi_{x_j}(x)+\tilde{z}_0\psi_0(x),
\label{eq:second_order_curve_constrained}
\end{equation}
with $\tilde{z}=\tilde{C}^\top\left(\tilde{C} \tilde{A} \tilde{C}^\top\right)^{-1} \tilde{p}$, $\tilde{C}=\mathrm{blkdiag}(C,1)$, 
$\tilde{p}=\begin{pmatrix}
p\\-r
\end{pmatrix}
$, $\tilde{\phi}(x)=\begin{pmatrix}
\phi(x)\\ \psi(x)
\end{pmatrix}$,
and $\tilde{A}$ is the positive definite $(N+1)\times (N+1)$ matrix with components
\[
\tilde{A}_{ij}=
\begin{cases}
A_{ij} & i\le j\le N\\
x_i	& i< j=N+1\\
1	& i=j=N+1\\
\end{cases}
.
\]

\section{Discount curve sensitivites}\label{sec:sensitivities}
The optimal discount curve \eqref{eq:second_order_curve} depends on the benchmark quotes through the vector $z=C^\top(CAC^\top)^{-1}p$. Depending on the type of benchmark instruments used, their quotes can enter through the price vector $p$ or through the cashflow matrix $C$. For example, prices of coupon bonds enter through $p$, while swap rates and forward rates enter through $C$. The results in this section are derived for the curve in \eqref{eq:second_order_curve}, however the results for the optimal curve \eqref{eq:second_order_curve_constrained} with constrained short rate directly follow by replacing $p,C,z,A,\phi$ with $\tilde{p},\tilde{C},\tilde{z},\tilde{A},\tilde{\phi}$, respectively.

\subsection{Portfolio hedging}\label{sec:hedging}
The sensitivities of the optimal discount curve $g^\ast(x;p,C)$ with respect to the entries of $p$ and $C$ are most easily expressed using directional derivatives:
\begin{lemma}\leavevmode
\begin{enumerate}
\item 
The directional derivative $D_p g^\ast\cdot v\in H$ of the optimal discount curve $g^\ast$ in \eqref{eq:second_order_curve} along a vector $v\in\R^n$ is given by
\begin{align*}
\left(D_p g^\ast\cdot v\right) (x) &= \sum_{i=1}^nv_i\frac{\partial g^\ast}{\partial p_i}(x)=c(v)^\top\phi(x),
%\sum_{j=1}^N c_j(v)\,\phi_{x_j}(x),
\end{align*} 
with $c(v)=C^\top\left(C A C^\top\right)^{-1} v\in\R^N$.
\item
The directional derivative $D_C g^\ast\cdot m\in H$ of the optimal discount curve $g^\ast$ in \eqref{eq:second_order_curve} along a matrix $m\in\R^{n\times N}$ is given by
\begin{align*} 
\left(D_C g^\ast \cdot m \right) (x)&= \sum_{i=1}^n\sum_{j=1}^N m_{ij}\frac{\partial g^\ast}{\partial C_{ij}}(x)=f(m)^\top \phi(x),
%\sum_{j=1}^N f_j(m)\,\phi_{x_j}(x),
\end{align*} 
with $f(m) =\left[ m^\top-C^\top(CAC^\top)^{-1}(CAm^\top+mAC^\top)\right](CAC^\top)^{-1}p\in\R^N$.
\end{enumerate}
\label{lemma:sensitivities}
\end{lemma}

These sensitivities can be used in practice to hedge a portfolio of securities against changes in the discount curve. Consider for example a bond portfolio which generates fixed cashflows $c_k$ in $\tau_k$ years, $\tau_k\in[0,\bar{\tau}]$, $k=1,\ldots,K$, and denote its current value by $V_{port}$. Suppose that all benchmark instruments are coupon bonds.\footnote{A similar hedging strategy can be built if the benchmark instruments have quotes that enter through $C$ using the directional derivative with respect to $C$ from Lemma \ref{lemma:sensitivities}.} A change $\Delta p_i$ in the price of the $i$-th benchmark instrument leads to the following change $\Delta V_{port}$ in the value of the bond portfolio:
\begin{align*}
\Delta V_{port}&=\sum_{i=1}^n\frac{\partial V_{port}}{\partial p_i}\Delta p_i= \sum_{i=1}^n \underset{=:q_i}{\underbrace{\sum_{k=1}^K c_k\left(D_p g^\ast\cdot e_i\right) (\tau_k)}}\Delta p_i,
\end{align*}
where $e_i\in\R^n$ denotes the $i$-th canonical basis vector. Hence, we can hedge the bond portfolio against changes in the prices of the benchmark coupon bonds by purchasing $-q_i$ units of the $i$-th benchmark coupon bond. \cite{andersen2007discount} points out that such a hedging strategy only works well in practice if the discount curve construction produces `local perturbations'. For example if bond $i$ has a short maturity, then $(D_p g^\ast\cdot e_i)(x)$ should be zero for large $x$ in order to avoid hedging long-term cashflows with short-term instruments. In general, cubic splines are known to perform poor with this respect and the above hedging strategy might therefore give unreasonable results with the discount curve construction presented in this paper. 

Hedging against individual small movements of the benchmark prices is however not necessarily consistent with the way interest rates move over time. Indeed, there is abundant empirical evidence that interest rate movements are attributable to a small number of stochastic factors often called level, slope, and curvature (see e.g., \cite{litterman1991common}). An alternative to hedging against changes in each benchmark price is therefore to directly hedge against interest rate movements that are deemed most likely. Specifically, we first build a discount curve $g^\ast(x)$ and then consider functional shifts $s_j(x)$, $j=1,\ldots,J$, to, for example, the corresponding forward curve $f^\ast(x)$. The sensitivity of $V_{port}=V_{port}(f^\ast)$ to these functional shifts can be expressed through the following functional derivative:
\[
\frac{\dd V_{port}(f^\ast+\epsilon s_j)}{\dd \epsilon}\Bigg\vert_{\epsilon =0}=-\sum_{k=1}^K c_k g^\ast(\tau_k)\int_0^{\tau_k}s_j(x)\,\dd x,\quad j=1,\ldots J.
\]
Next, we construct a hedging portfolio such that the functional derivatives of the hedged portfolio are equal to (or as close as possible to) zero.\footnote{We refer to section 6.4.2-6.4.3 in \cite{andersen2010interest} for more details on this hedging approach.} 
%directly perturb the corresponding forward curve $f^\ast$:
%\[
%f_{j}(x):=f^\ast(x)+s_j(x),\quad 0\le x\le \bar{\tau},\quad j=1,\ldots, J,
%\]
%where $s_j$ represent small perturbations of the forward curve. Next we value the portfolio to be hedged using the perturbed curves $f_j$ and determine the hedging portfolio needed to offset the portfolio value changes caused by the perturbations.\footnote{We refer to section 6.4.2-6.4.3 in \cite{andersen2010interest} for more details on this hedging approach.} 
The main advantage of this approach is that the method used to construct $g^\ast$ does not play a major role in determining the hedging strategies (see e.g., \cite{hagan2006interpolation}). If $J=1$ and $s_1(x)\equiv 1$, then we have a standard duration hedge. In order to hedge more than only parallel shifts in the forward curve, a popular choice in practice for $s_j$ are piecewise triangular functions around a pre-defined set of so called key rate horizons $0\le \xi_1<\dots< \xi_J\le \bar{\tau}$:
 \begin{align*}
&s_1(x)=
\begin{cases} 
\frac{\xi_{2}-x}{\xi_2-\xi_1} & x\in [\xi_1,\xi_2]\\
0 & else
\end{cases},\quad
s_J(x)=
\begin{cases} 
\frac{x-\xi_{J-1}}{\xi_J-\xi_{J-1}} & x\in [\xi_{J-1},\xi_J]\\
0 & else
\end{cases}\\
&s_j(x)=
\begin{cases} 
\frac{x-\xi_{j-1}}{\xi_j-\xi_{j-1}} & x\in [\xi_{j-1},\xi_j]\\
\frac{\xi_{j+1}-x}{\xi_{j+1}-\xi_{j}} & x\in [\xi_{j},\xi_{j+1}]\\
0& else
\end{cases},\quad j=2,\ldots, J-1.
 \end{align*}

\subsection{Optimal market quotes} \label{section:sec4}
So far we have assumed that market quotes are observed without any error. In practice, however, we do not observe a single price but rather a bid-ask range. Any price in this range can be used to estimate the discount curve and this flexibility can be used to increase the smoothness of the discount curve. 

The following lemma provides an explicit expression for the norm of the optimal discount curve we have derived before (i.e., for the case with equality constraints):
\begin{lemma}\label{lemma_optimal_norm}
The squared norm of the optimal discount curve $g^\ast$ in \eqref{eq:second_order_curve} is given by
\[
\Vert g^\ast \Vert^2_H=p^\top (CAC^\top)^{-1}p.
\]
\end{lemma}

We first assume that the benchmark instruments are coupon bonds for which we observe bid prices $p_b\in\R^n$ and ask prices $p_a\in\R^n$. We are now interested in solving the following optimization problem:
\begin{equation}
\begin{array}{cc}
\displaystyle
\min_{p\in \R^n} & p^\top (CAC^\top)^{-1}p\\
\text{s.t.}			 &p_b\leq p\leq p_a
\end{array}.
\label{optim4}
\end{equation}
Remark that $(CAC^\top)^{-1}\in\R^{n\times n}$ is a positive definite matrix. Indeed, $C$ is assumed to have full rank and $A$ is positive definite as a consequence of the definiteness of the inner product. Hence, \eqref{optim4} is a convex quadratic programming problem where the unique solution $p^\ast$ can easily be found using standard techniques. The optimal discount curve is then given by $g^\ast=M^+p^\ast$.

Assume now that the benchmark instruments have quotes that enter through $C$. This is for example the case for swaps and FRAs. Denote the bid and ask quotes by $\alpha_b$ and $\alpha_a$, respectively. The optimization problem now becomes:
\begin{equation}
\begin{array}{cc}
\displaystyle
\min_{\alpha\in \R^n} & p^\top (CAC^\top)^{-1}p\\
\text{s.t.}			 &\alpha_b\leq \alpha\leq \alpha_a
\end{array}.
\label{optim5}
\end{equation}
This problem is more difficult to solve than \eqref{optim4} because it is not necessarily a convex programming problem. However, we are able to compute the gradient explicitly:
\begin{lemma} The partial derivative with respect to $\alpha_i$ of the squared norm of the optimal discount curve in \eqref{eq:second_order_curve} is given by:
\[
\frac{\partial \Vert g^\ast\Vert_H^2}{\partial \alpha_i}= -2p^\top (CAC^\top)^{-1}\frac{\partial C}{\partial \alpha_i}AC^\top (CAC^\top)^{-1}p,
\]
where $\frac{\partial C}{\partial \alpha_i}$ denotes the componentwise derivative of $C$ with respect to the quote $\alpha_i$.
\end{lemma}
We can therefore use a wide range of gradient-based constrained optimization algorithms. 
A similar idea was used by \citet{kwon2002general}, however his approach requires a numerical evaluation of the gradient at every iteration step. In contrast, we have the gradient in closed form which can be beneficial for both the computation time and accuracy of the numerical procedure.

\section{Numerical examples}
In this section we discuss three practical illustrations of the pseudoinverse method using different types of benchmark instruments.
\label{sec:examples}
\subsection{Coupon bonds}
%\normalsize
In this example we estimate the discount curve from prices of coupon bonds. Specifically, we consider data from 4th of September 1996 on nine UK government bonds with semi-annual coupons and times to maturity varying approximately from 2 months to 12 years, see Table \ref{table:gilt} for details. The vector $p$ and the first ten columns (out of 104+1) of the matrix $C$ are shown below:

\scriptsize
\[ 
p
=
\left(
\begin{array}{c}
1\\
103.82\\
106.04\\
118.44\\
106.28\\
101.15\\
111.06\\
106.24\\
98.49\\
110.87
\end{array}\right),\quad
C=\left(
\begin{array}{cccccccccccc}
1& 0 & 0 & 0 & 0 & 0 & 0 & 0 & 0 &  0 &0& \dots \\
0 &0 & 0 & 0 & 105 & 0 & 0 & 0 & 0 & 0 &  0 & \dots \\
0 &0 & 0 & 0 & 0 & 0 & 4.875 & 0 & 0 & 0 & 0 & \dots \\
0 &6.125 & 0 & 0 & 0 & 0 & 0& 0 & 0 & 0 & 6.125 & \dots \\
0 &0 & 0 & 0 & 0 & 0 & 0 & 0 &  4.5 & 0& 0 & \dots \\
0 &0 & 0 & 3.5 & 0 & 0 & 0 & 0 & 0 & 0 & 0 & \dots \\
0 &0 & 0 & 0 & 0 & 0 & 0 & 4.875 & 0 & 0 & 0 & \dots \\
0 &0 & 0 & 0 & 0 & 4.25 & 0 & 0 & 0 & 0 & 0 & \dots \\
0 &0 & 0 & 0 & 0 & 0 & 0 & 0 & 0 & 3.875 & 0 & \dots \\
0 &0 & 4.5 & 0 & 0 & 0 & 0& 0 & 0 & 0 & 0 & \dots
\end{array}\right).\]
\normalsize

\begin{figure}
\begin{subfigure}{0.5\textwidth}
\center
\includegraphics[width=\textwidth]{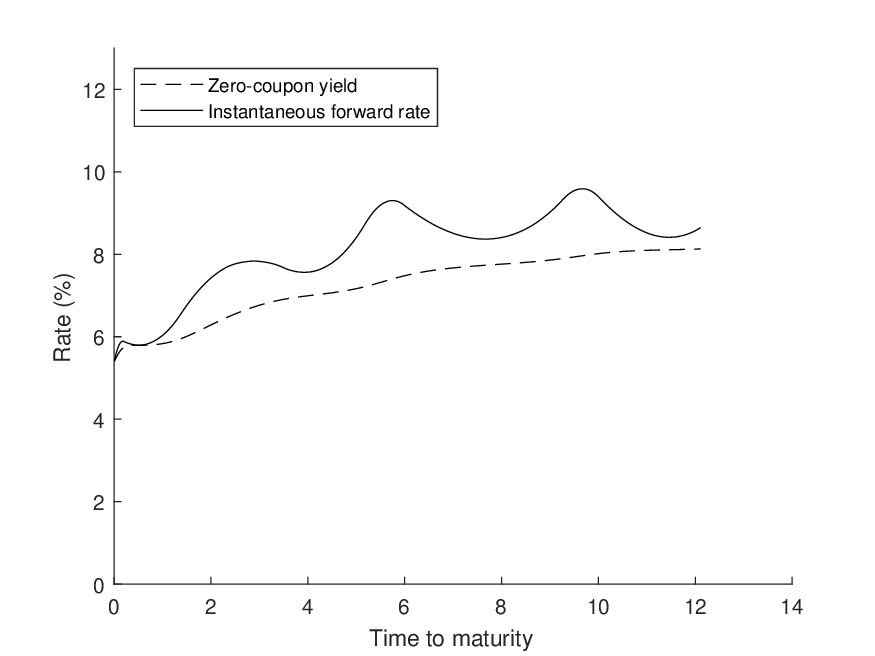}
\caption{Original prices.}
\label{fig:gilt_orig}
\end{subfigure}
\begin{subfigure}{0.5\textwidth}
\center
\includegraphics[width=\textwidth]{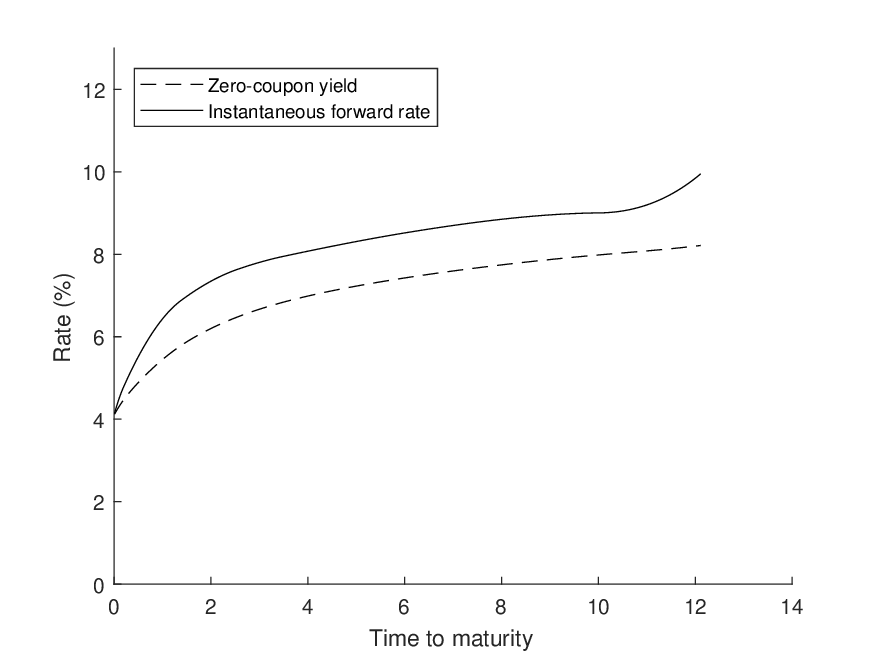}
\caption{Optimal prices}
\label{fig:gilt_optim}
\end{subfigure}

\begin{subfigure}{0.5\textwidth}
\center
\includegraphics[width=\textwidth]{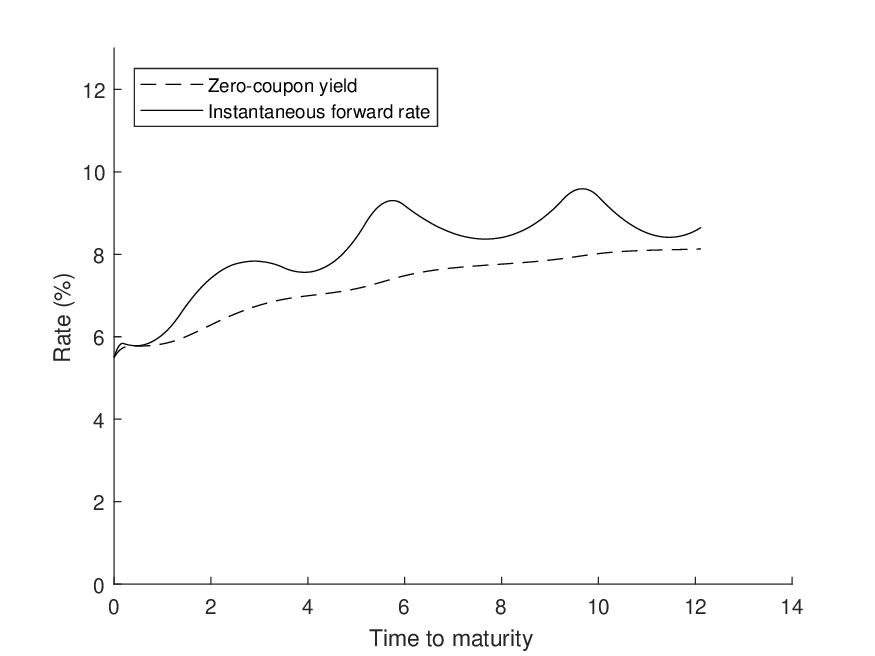}
\caption{Original prices, fixed $r=5.50\%$.}
\label{fig:gilt_orig_fixed_short}
\end{subfigure}
\begin{subfigure}{0.5\textwidth}
\center
\includegraphics[width=\textwidth]{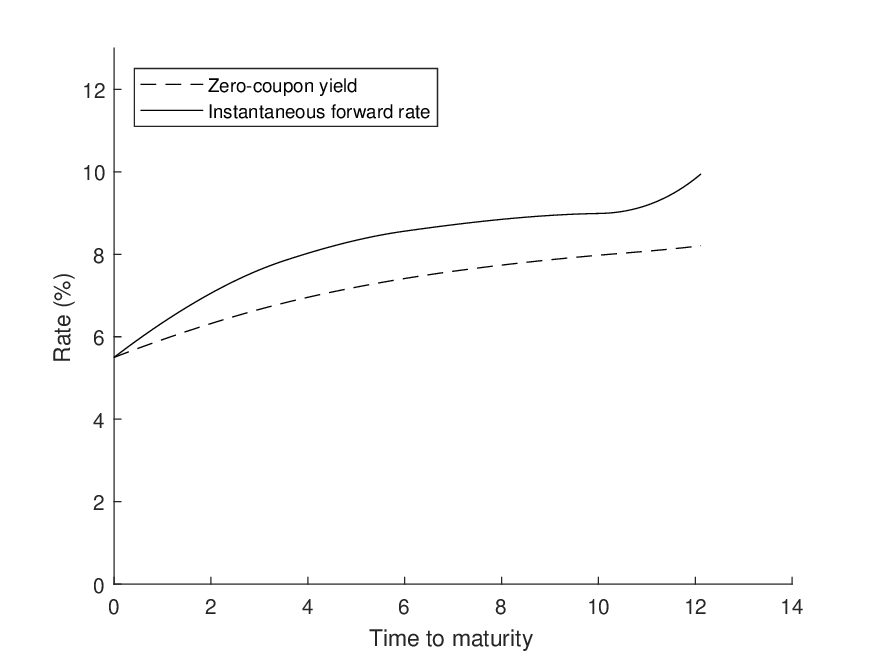}
\caption{Optimal prices, fixed $r=5.50\%$.}
\label{fig:gilt_optim_fixed_short}
\end{subfigure}
\caption{Zero-coupon yield and instantaneous forward rate from the discount curve estimated using prices of UK government bonds.}
\label{fig:gilt}
\end{figure}

The first row and column of $C$ and $p$ correspond to the restriction $g(0)=1$. Figure \ref{fig:gilt_orig} shows the continuously compounded yield curve and the instantaneous forward curve obtained from the pseudoinverse method presented in Section \ref{sec:theory}. The yield curve looks very smooth, but the forward curve exhibits oscillatory behavior that might be undesirable. 

Next, we assume that the observed coupon bond prices are mid prices and we assume a relative bid-ask spread of $0.50\%$ for every bond price. We use the approach outlined in Section \ref{section:sec4} to find a discount curve which produces coupon bond prices within the bid-ask range. Figure \ref{fig:gilt_optim} shows the resulting yield curve and instantaneous forward curve. We observe a significant increase in smoothness for the forward curve. Remark also the decrease of the short rate from approximately $5.5\%$ to $4\%$ as a consequence of the $g'(0)^2$ term in the norm definition \eqref{eq:norm2}. As explained at the end of Section \ref{sec:theory} this can be avoided by fixing the short rate to an exogenous constant, for example $r=5.5\%$. The result of this estimation is shown in Figure \ref{fig:gilt_orig_fixed_short} and Figure \ref{fig:gilt_optim_fixed_short} for the original and optimal prices, respectively.

\subsection{Libor single curve}\label{secSCE}
In this example we use the same curve for both discounting cashflows as well as projecting forward rates. We consider data from the US money and swap markets as of 1st of October 2012, as shown in Table~\ref{tab:quotes_single_curve}. More specifically we look at three USD Libor rates (overnight, 1M and 3M) with maturity dates $S=\{S_1,S_2,S_3\}$, five futures contracts on the 3M Libor and nine par swap rates with annual paying fixed leg. The futures contracts are quoted as:
\[
100(1-F_{futures}(T_{i-1},T_i)),\quad i=1,\ldots,7,
\]
with $F_{futures}(T_{i-1},T_i)$ denoting the futures rate and $T=\{T_0,T_1,\ldots,T_7\}$ the corresponding reset/settlement dates. We ignore any convexity adjustments and take the futures rate as the simple forward rate to keep the estimation procedure model-independent.\footnote{The convexity adjustments are only a fraction of basis points because of the short maturities, so they do not make much qualitative difference.} Finally we denote by $U=\{U_1,\ldots,U_{30}\}$ the cashflow dates of the swaps.

\subsubsection*{Traditional bootstrap}
We first perform a traditional bootstrap where we interpolate all the missing simple spot and swap rates linearly. Remark first the overlapping cashflow dates of the different instruments:
\begin{equation}
S_1<S_2<T_0<S_3<T_1<T_2<T_3<U_1<T_4<T_5<U_2<\cdots<U_{30}.
\label{eq:cashflow_dates}
\end{equation}
The prices of the discount bonds $g(S_1)$, $g(S_2)$, and $g(S_3)$ are readily obtained from the given Libor rates. At the reset date of the first futures contract, we obtain the simple spot rate  $L(T_0)$ by linear interpolation of the last two Libor rates:
\[
L(T_0)=wL(S_2)+(1-w)L(S_3), \quad w=\frac{\delta(T_0,S_3)}{\delta(S_2,S_3)},
\]
where $\delta(x,y)$ denotes the day count fraction between dates $x$ and $y$.
The discount factor $g(T_0)$ is now recovered from the interpolated simple spot rate $L(T_0)$. Treating the futures rates as simple forward rates, we have all the information needed to compute $g(T_i)$, $i=1,\ldots,5$, iteratively from:
\[
P(T_i)=\frac{g(T_{i-1})}{1+\delta(T_{i-1},T_i)F(T_{i-1},T_i)}.
\]
For the swaps we exploit again the overlapping cashflow dates to obtain $g(U_1)$ by linearly interpolating between $L(T_3)$ and $L(T_4)$. The swap rate $R_{swap}(U_1)$ of the swap with just one cashflow at $U_1$ can be calculated from the discount bond price $g(U_1)$. The remaining discount factors are now obtained by iterated use of the formula:
\[
g(U_i)=\frac{1-R_{swap}(U_i)\sum_{j=1}^{i-1}\delta(U_{j-1},U_j)g(U_j)}{1+R_{swap}(U_i)\delta(U_{i-1},U_i)},\quad i=2,\ldots,30,
\]
where all the missing swap rates are obtained by linear interpolation. 

Finally, we obtain the discount curve for the continuum of maturities by linearly interpolating the zero-coupon bond yields between cashflow dates. Figure \ref{fig:bootstrap_long} and Figure \ref{fig:bootstrap_short} show the zero-coupon bond yields and the instantaneous forward rates\footnote{We have approximated the instantaneous forward rates using first order finite differences on a fine grid.}. Although the zero-coupon yield curve looks fairly smooth, the same cannot be said of the instantaneous forward curve. This well known `sawtooth'-behaviour of the forward curve is a consequence of the linear interpolation. Other interpolation techniques may lead to improved smoothness of the forward curve, however choosing the correct technique remains somewhat arbitrary and can lead to a significant increase in complexity.

\subsubsection*{Pseudoinverse}
The pseudoinverse method that we introduced in Section~\ref{sec:theory} does not require any ad hoc interpolation. We only have to construct the cashflow matrix and the smoothness maximization criterion uses the remaining degrees of freedom in an optimal way. The cashflow matrix $C$ in this example has dimension $18\times 40$, one row for every observed instrument and an additional row to impose $g(0)=1$. The columns represent all the dates relevant in the valuation of the instruments.

The Libor rates $L(S_i)$, $i=1,2,3$, can be represented as instruments that have price $1$ today and cashflow $1+\delta(0,S_i)L(S_i)$ at time $S_i$. The simple forward rates $F(T_{i-1},T_i)$, $i=1,\ldots,5$, can be seen as instruments with price $0$ today, cashflow $-1$ at time $T_{i-1}$ and another cashflow of $1+\delta(T_{i-1},T_{i})F(T_{i-1},T_i)$ at time $T_i$. For the swaps with maturity $U_i$, $i=2,3,4,5,7,10,15,20,30$, we recall the definition of the par swap rate $R_{swap}(U_i)$:
\[
1-g(U_i)=R_{swap}(U_i)\sum_{j=1}^i\delta(U_{j-1},U_j) g(0,U_j)
\]
where we set $U_0:=0$. We see that this is equivalent to an instrument with price $1$ today, cashflow $\delta(U_{j-1},U_j)R_{swap}(U_i)$ at time $U_j$, $j=1,\ldots,i-1$, and a final cashflow $1+\delta(U_{i-1},U_i)R_{swap}(U_i)$ at time $U_i$. The vector $p$ and the first 13 columns of the matrix $C$ therefore take the following form:

\scriptsize
\[
p=
\begin{blockarray}{c}
\phantom{p}\\
\begin{block}{(c)}
1\\
1\\
1\\
1\\
0\\
0\\
0\\
0\\
0\\
1\\
1\\
1\\
1\\
1\\
1\\
1\\
1\\
1\\
\end{block}
\end{blockarray}
,\, C=
\begin{blockarray}{cccccccccccccc>{\scriptstyle}c}
0&S_1&S_2&T_0&S_3&T_1&T_2&T_3&U_1&T_4&T_5&U_2&U_{3}&\cdots\\
\begin{block}{(cccccccccccccc)>{\scriptstyle}c}
1&0 & 0 	& 0 &0&0&0&0&0&0&0&0&0&\cdots&g(0)=1\\
0&c_{11} & 0 	& 0 &0&0&0&0&0&0&0&0&0&\cdots&Libor\\
0&0		&c_{22}	&0&0&0&0&0&0&0&0&0&0&\cdots&Libor\\
0&0 		&  0 	&  	0 & c_{34} &  0 &   0& 0 &  0 &  0 &  0 &  0 &  0 &  \cdots&Libor\\
0&0 		&  0 	&	-1&  0 &  c_{45} &  0 &  0 & 0 &  0 &  0 &  0 &  0 &  \cdots&Futures\\
0&0 		&  0 	&  0 &  0 & -1 &  c_{56} &  0 &  0 &0 &  0 &  0 &  0 &  \cdots&Futures\\
0&0 		&  0 &  0 &  0 &  0 &  -1 & c_{67} &  0 &  0 &0 &  0 &  0 &  \cdots&Futures\\
0&0 		&  0 &  0 &  0 &  0 &  0 &  -1 &  0 & c_{79} &0 &  0 &  0 &  \cdots&Futures\\
0&0 		&  0 &  0 &  0 &  0 &  0 &  0 &  0 &   -1 & c_{8,10} & 0 &0 &  \cdots&Futures\\
0&0 		&  0 &  0 &  0 &  0 &  0 &  0  &  c_{98} &  0   &0 &  c_{9,11} &  0 &  \cdots&Swap\\
0&0 		&  0 &  0 &  0 &  0 &  0 &  0  &  c_{10,8} &  0   &0 &  c_{10,11} &  c_{10,12} &  \cdots&Swap\\
0&0 		&  0 &  0 &  0 &  0 &  0 &  0  &  c_{11,8} &  0   &0 &  c_{11,11} &  c_{11,12} &  \cdots&Swap\\
0&0 		&  0 &  0 &  0 &  0 &  0 &  0  &  c_{12,8} &  0   &0 &  c_{12,11} &  c_{12,12} &  \cdots&Swap\\
0&0		&  0 &  0 &  0 &  0 &  0 &  0  &  c_{13,8} &  0   &0 &  c_{13,11} &  c_{13,12} &  \cdots&Swap\\
0&0 		&  0 &  0 &  0 &  0 &  0 &  0  &  c_{14,8} &  0   &0 &  c_{14,11} &  c_{14,12} &  \cdots&Swap\\
0&0 		&  0 &  0 &  0 &  0 &  0 &  0  &  c_{15,8} &  0   &0 &  c_{15,11} &  c_{15,12} &  \cdots&Swap\\
0&0 		&  0 &  0 &  0 &  0 &  0 &  0  &  c_{16,8} &  0   &0 &  c_{16,11} &  c_{16,12} &  \cdots&Swap\\
0&0 		&  0 &  0 &  0 &  0 &  0 &  0  &  c_{17,8} &  0   &0 &  c_{17,11} &  c_{17,12} &  \cdots&Swap\\
\end{block}
\end{blockarray}.
\]
\normalsize

\begin{figure}
\begin{subfigure}{0.5\textwidth}
\center
\includegraphics[width=\textwidth]{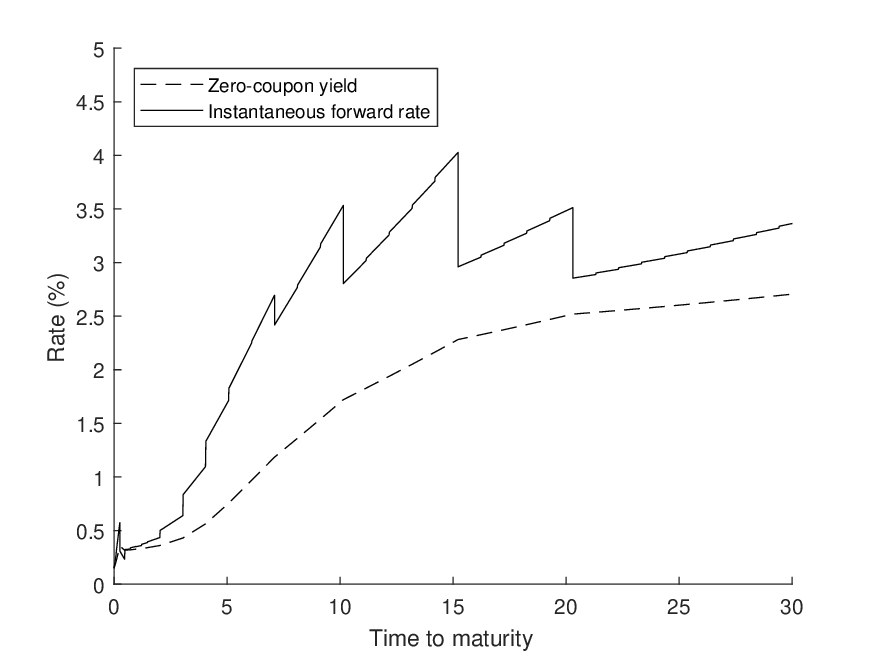}
\caption{Bootstrap.}
\label{fig:bootstrap_long}
\end{subfigure}
\begin{subfigure}{0.5\textwidth}
\center
\includegraphics[width=\textwidth]{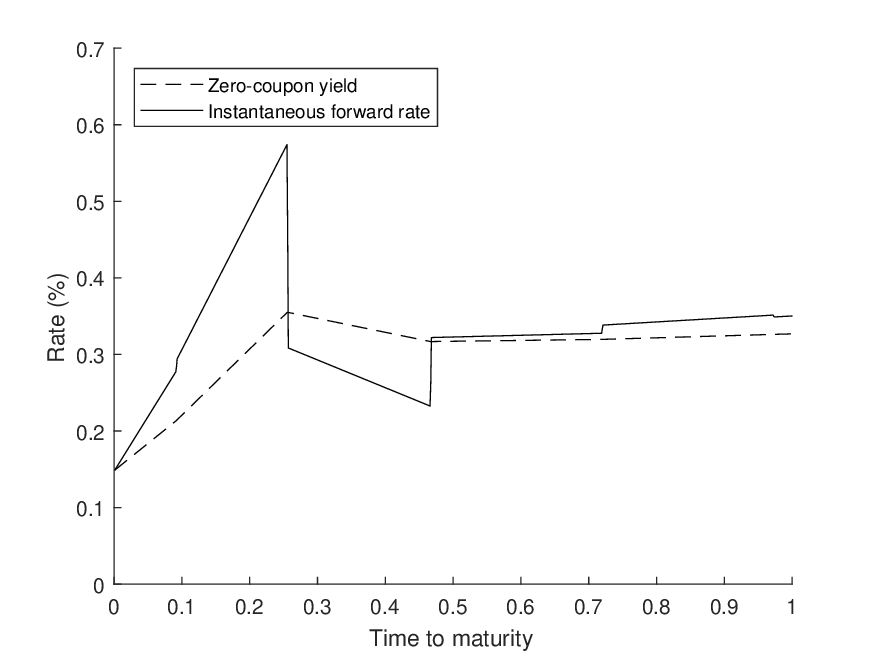}
\caption{Bootstrap, zoom.}
\label{fig:bootstrap_short}
\end{subfigure}
\begin{subfigure}{0.5\textwidth}
\center
\includegraphics[width=\textwidth]{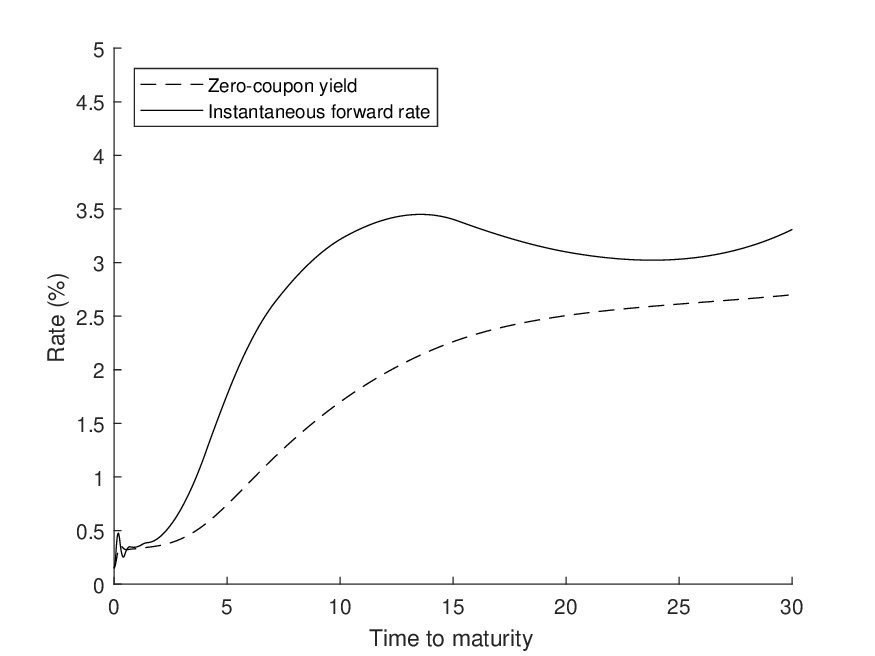}
\caption{Pseudoinverse.}
\label{fig:PI_interp_2}
\end{subfigure}
\begin{subfigure}{0.5\textwidth}
\center
\includegraphics[width=\textwidth]{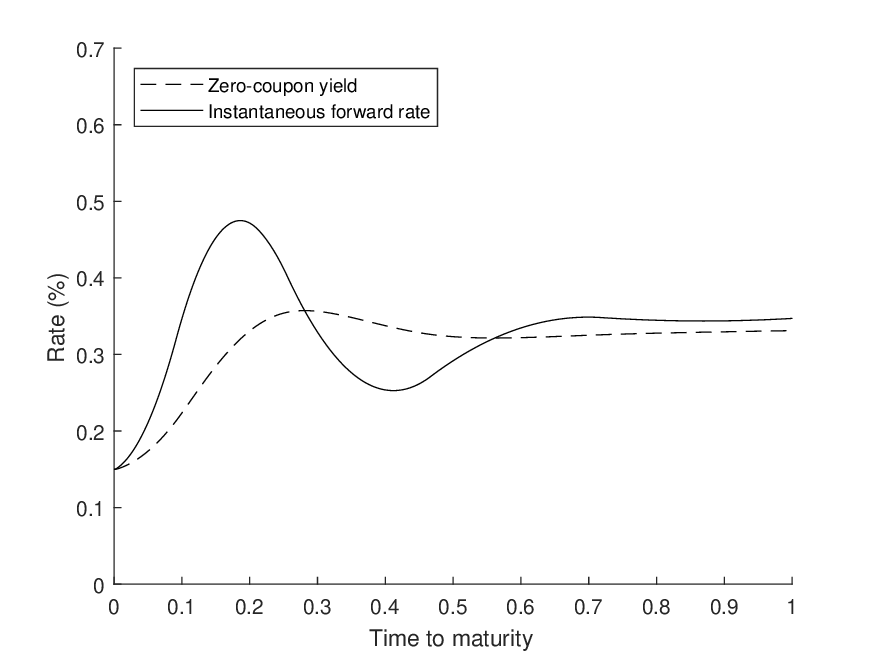}
\caption{Pseudoinverse, zoom.}
\label{fig:PI_interp_2_short}
\end{subfigure}
\caption{Zero-coupon yield and instantaneous forward rate from the discount curve estimated using Libor instruments.}
\label{fig:libor_single_curve}
\end{figure}

In Figure \ref{fig:PI_interp_2} and Figure \ref{fig:PI_interp_2_short} we have plotted zero-coupon yields and instantaneous forward rates. We observe that the pseudoinverse method produces a substantially smoother forward curve than the one in Figure \ref{fig:bootstrap_long}. Note that fixing the short rate to an exogenous constant would not change much here since we have included the overnight Libor rate as a benchmark instrument.

\begin{figure}
\center
\includegraphics[scale=0.8]{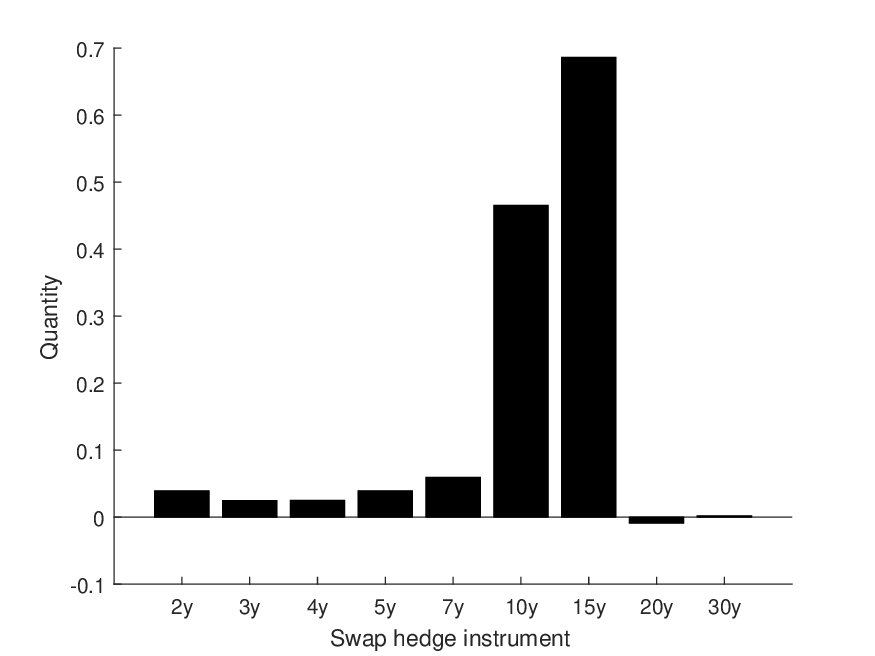}
\caption{Hedging a 13 year swap using the nine swaps used in the estimation as hedging instruments.}
\label{fig:hedge}
\end{figure}

As an example of a hedging strategy, we consider a payer swap maturing in 13 years. We want to hedge this swap with the nine receiver swaps used in our estimation through the hedging approach outlined in Section \ref{sec:hedging} with triangular functional shifts to the forward curve. Following \cite{andersen2010interest} we choose to use key rate horizons $\xi_j$, $j=1,\ldots,J$, spaced three months apart over the interval $[0,\bar{\tau}]$. Note that $J>9$, which means we cannot build a perfect hedge but only an approximate one (in a least squares sense). The hedging quantities are shown in Figure \ref{fig:hedge}. The hedging strategy consists roughly in combining the swaps with maturity in 10 and 15 years. This makes sense intuitively since these are the two hedging instruments whose cashflows resemble the one of the 13 year swap the most.

\subsection{Libor multi curve}\label{secMCE}
After the credit crisis of 2008 it became clear that using one and the same curve for both discounting and projecting cashflows was no longer realistic. Today's market standard is to use overnight indexed swaps (OIS) to extract a risk-free curve for the purpose of discounting cashflows and to use separate curves to project forward rates with different tenors. We show in this example how the pseudoinverse method can easily be used to extract all these different curves from market data.

Table \ref{table:multicurve} shows quotes of four different swap instruments from the eurozone market as of 4th of November 2013. The first is an OIS that pays fixed and receives floating tied to EONIA. EONIA swaps with maturity longer than one year have annual payment frequency on both legs while those with maturity less than one year only have a cashflow at maturity. The second swap instrument pays fixed and receives floating tied to 6M Euribor. The 6M Euribor swap has annual payments on its fixed leg and semi-annual on the floating leg. The remaining two swap instruments are basis swaps that swap cashflows tied to floating rates. The first one swaps 3M Euribor against 6M Euribor and the second one 1M Euribor against 6M Euribor. These swaps are quoted in terms of the spread that has to be added to the shorter leg such that the two legs have identical present value.

\begin{figure}
\center

\begin{subfigure}{0.8\textwidth}
\center
\includegraphics[width=\textwidth]{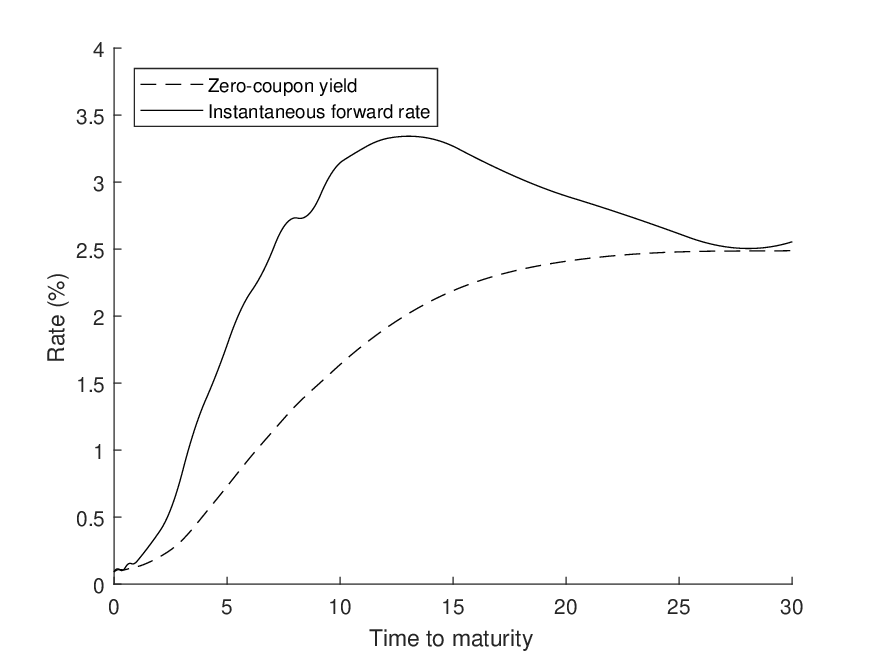}
\caption{Zero-coupon yield and instantaneous forward curve corresponding to the OIS discount curve $g_{OIS}$.}
\label{fig:ois}
\end{subfigure}
\begin{subfigure}{0.8\textwidth}
\center
\includegraphics[width=\textwidth]{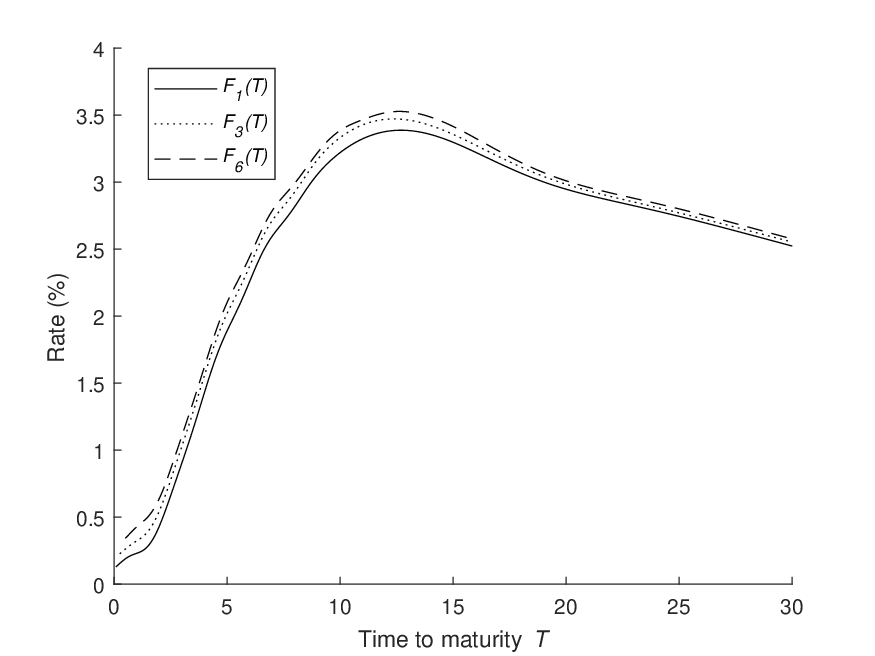}
\caption{Forward curves with tenors one, three, and six months.}
\label{fig:multicurve_forwards}
\end{subfigure}
\caption{Multicurve estimation with pseudoinverse method.}
\label{fig:multicurve}
\end{figure}

The curve $g_{OIS}$ used for discounting is extracted from the OIS quotes. This can be done with the pseudoinverse method in exactly the same way as in the single curve example. The corresponding yield and instantaneous forward curve are plotted in Figure \ref{fig:ois}. The 6M Euribor swap and the two basis swaps have payment frequencies that are multiples of one month. In the following we therefore consider a time grid $T=\{t_1,\ldots,t_{N}\}$ where $t_i$ and $t_{i-1}$, $i=1,\ldots,N$, are one month apart and $N=30\times 12$. 

We start with the instruments that swap the 6M Euribor against a fixed rate. These are quoted in terms of the swap rate $K$ that equates the value of the fixed and the floating leg:
\[
K\sum_{i=1}^n\delta(t_{12(i-1)},t_{12i})g_{OIS}(t_{12i})=\sum_{i=1}^{2n}\delta(t_{6(i-1)},t_{6i})g_{OIS}(t_{6i})F_{6}(t_{6i}),
\]
where $t_0=0$, $t_{12n}$ is the maturity of the swap, and $F_{k}(t_i)$ denotes the $k$-month simple forward rate with reference period $[t_{i-k},t_{i}]$. With our estimate for the OIS discount curve we are able to value the fixed leg of this swap. Notice that the right-hand side is a linear function of the unknown 6M forward rates with known coefficients. In other words, using our methodology from before we are able to extract the smoothest possible 6M forward curve that exactly fits the quoted swap rates. The pricing system therefore becomes $Cf=p$, where $f=(F_{6}(t_{6}),\ldots,F_{6}(t_{N}))^\top$, the price vector $p$ takes the form
\[
p=
\left(F_{6}(t_6),\,
K\delta(t_0,t_{12})g_{OIS}(t_{12}),\, \ldots,\,K\sum_{i=1}^{30}\delta(t_{12(i-1)},t_{12i})g_{OIS}(t_{12i})
\right)^\top
\]
and the first three rows of the `cashflow' matrix $C$ for become:

\scriptsize
\[
C=
\begin{pmatrix}
1 & 0 & 0 & 0 & 0 &  \cdots\\
\delta(t_{0},t_{6})g_{OIS}(t_{6})&\delta(t_{6},t_{12})g_{OIS}(t_{12})&0&0&0&\cdots\\
\delta(t_{0},t_{6})g_{OIS}(t_{6})&\delta(t_{6},t_{12})g_{OIS}(t_{12})&\delta(t_{12},t_{18})g_{OIS}(t_{18})&\delta(t_{18},t_{24})g_{OIS}(t_{24})&0&\cdots\\
\vdots&\vdots&\vdots&\vdots&\vdots&\ddots
\end{pmatrix}.
\]
\normalsize 

Next to the quotes of the swaps, we also included here the given 6M Euribor spot rate $F_{6}(t_6)$ by adding the row $(1,0,\ldots,0)$ to the matrix $C$ and the rate to the vector~$p$.

In a next step we use the 6M forward curve to extract the 3M forward curve from the 3M-6M tenor basis swaps. Basis swaps are quoted in terms of the spread $S$ that has to be added to the leg with the highest frequency in order to equate the value of both legs:
\begin{align*}
&\sum_{i=1}^{2n}\delta(t_{6(i-1)},t_{6i})g_{OIS}(t_{6i})F_{6}(t_{6i})\\
=&
\sum_{i=1}^{4n}\delta(t_{3(i-1)},t_{3i})g_{OIS}(t_{3i})(F_{3}(t_{3i})+S).
\end{align*}
Rearranging terms we get:
\begin{align*}
&\sum_{i=1}^{2n}\delta(t_{6(i-1)},t_{6i})g_{OIS}(t_{6i})F_{6}(t_{6i})-S\sum_{i=1}^{4n}\delta(t_{3(i-1)},t_{3i})g_{OIS}(t_{3i})\\
=&
\sum_{i=1}^{4n}\delta(t_{3(i-1)},t_{3i})g_{OIS}(t_{3i})F_{3}(t_{3i}).
\end{align*}
The left-hand side can be evaluated with the discount and 6M forward curve we have extracted earlier and the right-hand side is a linear function of the unknown 3M forward rates. We can therefore use the pseudoinverse method to extract the 3M forward curve. Once again we can also easily include the spot 3M Euribor rate $F_{3}(t_3)$. In a very similar fashion we also obtain the 1M forward curve from the 1M-3M basis swap quotes, where we also include the spot 1M Euribor rate $F_{1}(t_1)$. All three curves are plotted in Figure \ref{fig:multicurve_forwards}.

\begin{remark}
Although EONIA swaps are quoted up to 60 years of maturity, it is sometimes argued that the EONIA swaps with maturity longer than 1 year are not sufficiently liquid to be used in the construction of the OIS discount curve. The remaining part of the curve is instead often calculated from OIS-3M basis swaps, which are more actively traded. This does however create a circular dependency between the OIS discount curve, the 3M forward and 6M forward curves. The easy extension of the pseudoinverse method to the multi-curve world was mainly due to the fact that we can estimate all the curves one by one. If we have to estimate a part of the OIS discount curve from OIS-3M swaps, then we need to solve for all three curves at once. This increases the complexity of the problem because in the swap valuations the forward rates are multiplied with the discount rates, i.e.\ we face non-linear constraints in the optimization problem.
One possible workaround is to jointly estimate the OIS discount curve $g_1(x):=g_{OIS}(x)$ and the `discounted' forward curves $g_2(x):=g_{OIS}(x)F_{3}(x)$, $g_3(x):=g_{OIS}(x)F_{6}(x)$. By smoothing the discount curve and discounted forward curves, we are again solving an optimization problem with linear constraints:
\begin{equation}
\begin{array}{cc}
\displaystyle
\min_{g_1,g_2,g_3\in H} & \|g_1\|_{H}^2+\|g_2\|_{H}^2+\|g_3\|_{H}^2\\
\text{s.t.}			 & M_1g_1+M_2g_2+M_3g_3=p,
\end{array}
\nonumber
\end{equation}
for some appropriately defined linear maps $M_1$, $M_2$, and $M_3$. We leave the implementation of this extension for further research.
\end{remark}

\section{Pseudoinverse on the Euclidean space}\label{sec:finite_dim}
For readers unfamiliar with infinite-dimensional Hilbert spaces, we introduce in this section a finite-dimensional analogue of the method introduced in Section \ref{sec:theory}. Instead of looking for a discount curve in a Hilbert function space, we now search for a vector of discount factors in the Euclidian space which has maximal smoothness in some sense. Suppose we are interested in the discount factors at dates $0=u_1<\cdots<u_K$:
\[d=(g(u_1),\ldots,g(u_K))^\top,\]
for some $K\ge 1$ and (for simplicity) $u_{i+1}-u_i\equiv \delta>0$. Suppose furthermore that $d$ contains all the discount factors that are required to value the $n$ instruments we observe, i.e.\ $\{x_1,\ldots,x_N\}\subseteq \{u_1,\ldots,u_K\}$. Redefine $C\in\R^{n\times K}$ as the cashflow matrix of the $n$ benchmark instruments on the dates $\{u_1,\ldots,u_K\}$.

We cast the smoothness criterion \eqref{eq:norm2} in a discrete form using a left Riemann sum for the integral and forward finite differences for the derivatives (other choices are possible of course):
\begin{align*}
&g(0)^2+g'(0)^2+\int_{0}^{u_{K-1}}g''(x)^2\,\mathrm{d}x\\
&\approx g(u_0)^2+\frac{1}{\delta}(g(u_1)-g(u_0))^2+\sum_{i=0}^{K-2}\left(\frac{g(u_{i+2})-2g(u_{i+1})+g(u_i)}{\delta^2}\right)^2\delta\\
&=\lVert A d\rVert^2_K,
\end{align*}
where $\|\cdot\|_K^2$ denotes the Euclidian norm on $\R^K$ and 
\[
A=
\text{diag}(1,\delta^{-1/2},\delta^{-3/2},\ldots,\delta^{-3/2})
\scriptsize
\begin{pmatrix}
1&0&\dots&\dots&\dots&0\\
-1&1&0&&&\vdots\\
1&-2&1&\ddots&&\vdots\\
0&1&-2&1&\ddots&\vdots\\
\vdots&\ddots&\ddots&\ddots&\ddots&0\\
0&\dots&0&1&-2&1
\end{pmatrix}\in\R^{K\times K}.
\]

The finite dimensional optimization problem now becomes:
\begin{equation}
\begin{array}{rc}
\displaystyle
\min_{d\in \R^K} & \frac{1}{2}\|Ad\|_K^2\\
\text{s.t.}			 & Cd=p
\end{array}.
\label{eq:finite_dim}
\end{equation}
The above is a convex quadratic programming problem with linear inequality constraints, for which we obtain the following solution:
\begin{theorem}\label{thm:finite_dim}
There exists a unique solution $d^\ast\in \R^K$ to the optimization problem~\eqref{eq:finite_dim} and it is given as
\begin{equation*}
 d^\ast=A^{-1}M^+p,
\end{equation*}
where $M=CA^{-1}$ and $M^+=M^\top(M M^\top)^{-1}$ is the Moore--Penrose pseudoinverse of the matrix $M$.
\end{theorem}
Remark that in the finite-dimensional case it becomes very easy to impose positivity and monotonicity constraints on the discount factors:
\begin{equation*}
\begin{array}{rc}
\displaystyle
\min_{d\in \R^K} & \frac{1}{2}\|Ad\|_K^2\\
\text{s.t.}			 & Cd=p\\
& d_1>\cdots >d_K>0
\end{array}.
\end{equation*}
We therefore obtain a convex quadratic programming problem with linear inequality constraints. Such a problem has a unique solution that can easily be found with established algorithms implemented in many numerical software packages.

\section{Conclusion}\label{sec:conclusion}
We have introduced a novel method based on the Moore--Penrose pseudoinverse to extract a discount curve that exactly reproduces the prices of the benchmark instruments and that has maximal smoothness in the sense that it has minimal integrated squared second-order derivatives. The optimal discount curve is a piecewise-cubic function and is obtained as the unique solution of an infinite-dimensional optimization problem.
Bid-ask spreads can be incorporated to further increase the smoothness of the discount curve. The pseudoinverse method is very easy to implement, making it an interesting method of first resort before considering more complex alternatives.

\clearpage

\begin{appendix}

\section{Proofs}
This appendix contains all proofs.

\subsection*{Proof of Lemma \ref{lemma1}}
Integration by parts gives
\begin{align*}
g(\tau)&=g(0)+\int_0^\tau g'(x)\,\dd x\\
&=g(0)+\tau g'(0)-\int_0^\tau (x-\tau)g''(x)\,\dd x.
\end{align*}
From the definition of the scalar product $\langle\cdot,\cdot\rangle_{H}$ we get the following conditions for the function $\phi_\tau$:
\[
\left\{
\begin{array}{l}
\phi_\tau(0)=1\\
\phi'_\tau(0)=\tau\\
\phi''_\tau(x)=(\tau-x)1_{[0,\tau]}(x), \quad x\in [0,{\bar{\tau}}].
\end{array}
\right.
\]
Integrating two time we arrive at:
\[
\begin{array}{ll}
\phi'_\tau(x)&=\tau-\frac{1}{2}(x\wedge \tau)^2 +\tau(x\wedge\tau),\quad x\in [0,{\bar{\tau}}],\\
%\phi_\tau(x)&=1+\tau(x+\frac{1}{2}x^2)-\frac{1}{6}(x\wedge \tau)^3-\frac{\tau}{2}x^2+\frac{\tau}{2}(x\wedge \tau)^2+\frac{\tau^2}{2}(x-x\wedge \tau),\quad x\in [0,{\bar{\tau}}].\\
\phi_\tau(x)&=1-\frac{1}{6}(x\wedge \tau)^3+\frac{\tau}{2}(x\wedge \tau)^2-\frac{\tau^2}{2}(x\wedge \tau)+ x(1+\frac{\tau}{2})\tau,\quad x\in [0,{\bar{\tau}}].
\end{array}
\]
%The function $\phi_\tau$ is by construction three times differentiable over $[0,{\bar{\tau}}]$. In particular, it has an absolutely continuous first derivative and therefore $\phi_\tau\in H_2$

\subsection*{Proof of Theorem \ref{thm2}}
The transpose (adjoint operator) $M^\top:\R^n\to H$ of the linear map $M\colon H\to \R^n$ is defined by:
\[
\langle Mg,z\rangle_{\R^n}=\langle	g,M^\top z\rangle_{H},\quad \forall g\in H,\, \forall z\in\R^n.
\]
Using the definition of $M$ and the Riesz representation of the linear functional $\Phi$ we easily get:
\[ M^\top z =\sum_{j=1}^N  \phi_{x_j}C^\top_jz ,\quad z\in\R^n,\]
where $C_j$ represents the $j$-th column of the matrix $C$.

The Lagrangian $\mathcal{L}\colon H\times\R^n\to\R$ for problem \eqref{optim2} is defined as
\begin{align*}
\mathcal{L}(g,\lambda) &=\frac{1}{2} \|g\|^2_{H} +\lambda^\top\left( M g-p\right)\\
&=\frac{1}{2} \|g\|^2_{H}+\langle \lambda , Mg\rangle_{\R^n}-\langle \lambda , p\rangle_{\R^n}\\
&=\frac{1}{2} \|g\|^2_{H}+\langle M^\top\lambda , g\rangle_{H}-\langle \lambda , p\rangle_{\R^n}.
\end{align*}  
The optimizers $g^\ast$ and $\lambda^\ast$ satisfy the following first-order conditions with respect to the Fr\'echet derivative in $H$ and $\R^n$
\begin{align}
g^\ast+M^\top\lambda^\ast &=0 \label{eq:FOC_1}\\
M g^\ast-p&=0\label{eq:FOC_2}.
\end{align}
From \eqref{eq:FOC_1} we get $g^\ast=-M^\top\lambda^\ast$. Plugging this into \eqref{eq:FOC_2} gives $-M M^\top\lambda^\ast=p$. 
Observe now that $MM^\top:\R^n\to\R^n$ is a linear map that can be represented by the matrix $CAC^\top$, where $A$ is the positive definite $N\times N$ matrix with components
\begin{align*}
A_{ij} &= \langle \phi_{x_i},\phi_{x_j}\rangle_{H}=\phi_{x_i}(x_j)=\phi_{x_j}(x_i).
\end{align*}
We now obtain the following unique solution for the optimal Lagrange multiplier and optimal discount curve:
\[
\lambda^\ast=-\left(CAC^\top\right)^{-1}p,\quad g^\ast=M^\top \left(CAC^\top\right)^{-1}p.
\]
Note that the matrix $CAC^\top$ is invertible because $A$ is positive definite and $C$ has full rank.

The map
\[M^+\colon \R^n\to H,\, z\mapsto M^\top \left(M M^\top\right)^{-1}z,\] is known as the \emph{Moore--Penrose pseudoinverse} of the linear map $M$. We can therefore write the optimal discount curve as
\[
g^\ast = M^+ p.
\]

\subsection*{Proof of Lemma \ref{lemma:sensitivities}}
The optimal discount curve $g^\ast$ can be written as
\begin{equation}
 g^\ast(x) = \sum_{j=1}^N z_j\,\phi_{x_j}(x)=p^\top\left(C A C^\top\right)^{-1} C \phi(x),
 \label{proof3}
\end{equation}
with $\phi(x):=(\phi_{x_1}(x),\ldots,\phi_{x_N}(x))^\top$. Differentiating \eqref{proof3} with respect to the components of $p$ immediately gives the first statement of the theorem:
\[
(D_p g^\ast \cdot v)(x)=v^\top\left(C A C^\top\right)^{-1} C \phi(x).
\]
Differentiating \eqref{proof3} with respect to the cashflow $C_{ij}$ gives
\begin{align*}
\frac{\partial g^\ast}{\partial C_{ij}}(x)&= -p^\top \left(C A C^\top\right)^{-1} \frac{\partial C A C^\top}{\partial C_{ij}}\left(C A C^\top\right)^{-1} C\phi(x)+p^\top \left(C A C^\top\right)^{-1} I_{ij}\phi(x)\\
&=p^\top \left(C A C^\top\right)^{-1} \left(I_{ij}-(CAI_{ji}+I_{ij}AC^\top)\left(C A C^\top\right)^{-1} C\right)\phi(x),
\end{align*}
where $I_{ij}\in\R^{n\times N}$ denotes a matrix with the $(i,j)$-th entry equal to one and all of the other entries equal to zero. The second statement of the theorem now easily follows from the distributive property of matrix multiplication and
\[
\sum_{\substack{1\le i\le n\\1\le j\le N}}m_{ij}I_{ij}=m.
\]

\subsection*{Proof of Lemma \ref{lemma_optimal_norm}}
Using the notation of the proof of Theorem \ref{thm2}, the optimal discount curve $g^\ast$ can be written as
\[
g^\ast=M^+p=M^\top(MM^\top)^{-1}p.
\]
Using the fact that $M^\top$ is the dual operator of $M$, we get:
\begin{align*}
\Vert g^\ast\Vert^2 &=\langle g^\ast,g^\ast\rangle_H\\
&=\left\langle M^\top(MM^\top)^{-1}p,M^\top(MM^\top)^{-1}p\right\rangle_H\\
&=\left\langle (MM^\top)^{-1}p,MM^\top(MM^\top)^{-1}p\right\rangle_{\R^n}\\
&=\left\langle (CAC)^{-1}p,p\right\rangle_{\R^n}\\
&=p^\top(CAC^\top)^{-1}p.
\end{align*}

\subsection*{Proof of Theorem \ref{thm:finite_dim}}
The Lagrangian is defined as
\[
\Lcal(\lambda,d)=\frac{1}{2}\|Ad\|_K^2+\lambda^\top(Cd-p).
\]
The first-order optimiality conditions for the optimal $\lambda^\ast$ and $d^\ast$ are
\[
A^\top Ad^\ast+C^\top\lambda^\ast=0,\quad Cd^\ast-p=0.
\]
Straightforward calculations give the following unique solution:
\[
d^\ast=(A^\top A)^{-1}C^\top\Big(C(A^\top A)^{-1}C^\top\Big)^{-1}p.
\]
Using the fact that $A$ is invertible and defining $M:=CA^{-1}$, we finally obtain
\[
d^\ast=A^{-1}M^+p,
\]
where $M^+=M^\top(M M^\top)^{-1}$ is the Moore--Penrose pseudoinverse of the matrix $M$.

%%%%%%%%%%%%%%%%%%%%%%%%%%%%%%%%
%% TABLES
%%%%%%%%%%%%%%%%%%%%%%%%%%%%%%%%

\clearpage

\begin{table}
\begin{center}
\begin{tabular}{  l | r | c | c | r }
\hline
     & Coupon  & Next & Maturity &  Dirty price\\
     & (\%)    & coupon & date & ($p_i$) \\
     \hline
Bond 1 & 10 & 15/11/96 & 15/11/96 & 103.82 \\
Bond 2 & 9.75 & 19/01/97 & 19/01/98 & 106.04\\
Bond 3 & 12.25 & 26/09/96 & 26/03/99 & 118.44 \\
Bond 4 & 9 & 03/03/97 & 03/03/00 & 106.28\\
Bond 5 & 7 & 06/11/96 & 06/11/01 & 101.15\\
Bond 6 & 9.75 & 27/02/97 & 27/08/02 & 111.06\\
Bond 7 & 8.5 & 07/12/96 & 07/12/05 & 106.24 \\
Bond 8 & 7.75 & 08/03/97 & 08/09/06 & 98.49\\
Bond 9 & 9 & 13/10/96 & 13/10/08 & 110.87\\ \hline
\end{tabular}
\end{center}
\caption{Market prices for UK gilts, 04/09/1996.  Source: \cite{james2000interest}.}
\label{table:gilt}
\end{table}

\begin{table}
\center
\begin{tabular}{rrc}\toprule
Maturity date & Market quote (\%) & Instrument\\\midrule
$S_1=$ 04/10/2012	&0.1501& o/n Libor\\
$S_2=$ 05/11/2012	&0.2135& 1M Libor\\
$S_3=$ 03/01/2013	&0.3553&3M	Libor\\
$T_1=$ 20/03/2013	&99.685&	Futures\\
$T_2=$ 19/06/2013	&99.675&	Futures\\
$T_3=$ 18/09/2013	&99.655&	Futures\\
$T_4=$ 18/12/2013	&99.645&	Futures\\
$T_5=$ 19/03/2014	&99.620&	Futures\\
$U_2=$ 03/10/2014	&0.361&	Swap\\
$U_3=$ 05/10/2015	&0.431&	Swap\\
$U_4=$ 03/10/2016	&0.564&	Swap\\
$U_5=$ 03/10/2017	&0.754&	Swap\\
$U_7=$ 03/10/2019	&1.174&	Swap\\
$U_{10}=$ 03/10/2022	&1.683&	Swap\\
$U_{15}=$ 04/10/2027	&2.192&	Swap\\
$U_{20}=$ 04/10/2032	&2.405&	Swap\\
$U_{30}=$ 03/10/2042	&2.579&	Swap\\\bottomrule
\end{tabular}
\caption{Market quotes for Libor rates, futures prices and swap rates from the USD market as of 1st of October 2012. All contracts are spot ($T+2$) starting. Source: Bloomberg.}
\label{tab:quotes_single_curve}
\end{table}

\begin{table}
\center
\begin{tabular}{rrrrrr}
\toprule
\centering
Tenor	& EONIA-Fix (\%) &	6M-Fix (\%)&	3M-6M (bps)&	1M-3M (bps)& Cash (\%) \tabularnewline \midrule
o/n	&0.092&&&&\\		
1m	&0.102	&&&&0.129\\			
3m	&0.109	&&&&0.227\\			
6m	&0.108	&&&&0.342\\			
9m	&0.121	&&&&\\			
1y	&0.130	&0.386	&10.85	&8.90&\\
2y	&0.205	&0.482	&12.15	&10.90&\\
3y	&0.334	&0.656	&12.85	&12.60&\\
4y	&0.533	&0.870	&13.00	&13.70&\\
5y	&0.742	&1.097	&12.95	&14.20&\\
6y	&0.952	&1.306	&12.70	&14.30&\\
7y	&1.145	&1.503	&12.35	&14.20&\\
8y	&1.328	&1.677	&11.90	&14.00&\\
9y	&1.479	&1.833	&11.40	&13.80&\\
10y	&1.625	&1.973	&10.90	&13.60&\\
11y	&1.757	&2.095	&&&	\\
12y	&1.872	&2.199	&&&	\\
15y	&2.124	&2.418	&8.85	&11.80&\\
20y	&2.317	&2.570	&7.40	&10.10&\\
25y	&2.385	&2.618	&6.50	&8.90&\\
30y	&2.406	&2.625	&5.90	&8.10&\\\bottomrule
\end{tabular}
\caption{Mid swap rates for EONIA swaps, 6M Euribor swaps, 3M-6M basis swaps, 1M-6M basis swaps and Euribor cash rates as of 04/11/2013. Source: Bloomberg.}
\label{table:multicurve}
\end{table}

\newpage
\clearpage

%%%%%%%%%%%%%%%%%%%%%%%%%%%%%%%%
%% FIGURES
%%%%%%%%%%%%%%%%%%%%%%%%%%%%%%%%

\clearpage

\end{appendix}

\clearpage
%\printbibliography
\bibliography{references}
\bibliographystyle{chicago}

\end{document}